\documentclass[aps,prl,twocolumn, showpacs]{revtex4}

\usepackage{epsfig}
\usepackage{graphicx}

\def\be{\begin{equation}}
\def\ee{\end{equation}}
\def\bea{\begin{eqnarray}}
\def\eea{\end{eqnarray}}

\begin{document}

\title{Side jump contribution to spin-orbit mediated Hall effects and Berry curvature}
\author{Emmanuel I. Rashba}
\affiliation{Department of Physics and Center for Nanoscale Systems, Harvard University, Cambridge, Massachusetts 02138, USA\\ 
and Department of Physics, Loughborough University, Leicestershire LE11 3TU, UK}  
\date{\today}

\begin{abstract}
Anomalous Hall effect and spin Hall effect originate due to spin-orbit coupling that in the Kohn-Luttinger ${\bf k}\cdot{\bf p}$ formalism is represented by anomalous terms in the coordinate and velocity operators. Relation of these operators to the Berry curvature in the momentum space is presented for electrons in GaAs type semiconductors. For centrosymmetric semiconductors, transformational properties of Berry curvature are discussed.
\end{abstract}
\pacs{71.70.Ej, 72.10.Fk, 72.15.Gd,, 72.25.-b}

\maketitle

\narrowtext

Current research in semiconductor spintronics is driven by scientific curiosity and attractive perspectives of technological applications \cite{Wolf,ALS02,ZFDS04,Jake,HansonRMP,AwFl}. During the last decade, the field witnessed brilliant experimental achievements, enjoyed advancing new theoretical concepts and deeper understanding of underlying physical mechanisms, and demonstrated persistent efforts directed for applications. This recent progress is essentially based on prior work on spin-orbit coupling in crystals and electron confinement in nanoscale devices \cite{Been91}. An outstanding contribution into the early work on spin-orbit phenomena was made by experimental and theoretical work performed at the A. F. Ioffe Institute (St. Petersburg) since early 1970s. In the context of present paper, I emphasize prediction of the spin Hall effect (SHE) by D'yakonov and Perel' \cite{DP71} and of the direct and inverse photogalvanic effects by Ivchenko and Pikus \cite{IvPik78}. Both deal with generation of spin polarization by electric current, but first one near the edges while second one in the bulk.

Theory of spin-orbit phenomena in solids developed originally along a number of different lines related to the anomalous Hall effect (AHE) \cite{KarLut}, electric dipole spin resonance (EDSR) \cite{R60,RS91}, and optical orientation \cite{OptOr}. However, while the fields matured, close connections between them became more visible. E.g., identical equations for spin scattering probabilities describe the photogalvanic effect \cite{BIS82} and AHE \cite{SNMD06}, as noticed recently by Sinitsyn \cite{Sin08}. Similarly, identical mechanisms underlie the collisional narrowing of EDSR lines \cite{MelR} and the D'yakonov-Perel' spin relaxation \cite{DPrel71}, as is evident from a recent study by Duckheim and Loss \cite{DL06}. Also, there exists an intimate connection between the AHE and SHE \cite{Sin08,Schl06,ERH07}, therefore, they confront the same challenges that are discussed below. Meanwhile, the difference between them is rather essential. AHE is a bulk effect because of electric current conservation, while SHE critically depends also on boundary conditions because of spin current nonconcervation both in the bulk and at the edge. E.g., SHE is possible even in media where bulk spin currents (a somewhat ambiguous notion) vanish identically \cite{Rphys06,AB05,GBDS06,THKB07}. If to add spin interference effects initiated by Datta and Das \cite{DatDas}, spin blockade in strongly confined systems \cite{Jake,HansonRMP}, and giant magnetoresistance \cite{GMR1,GMR2}, all these fields of research merged gradually into a single body of {\it solid state spintronics.}

Anomalous Hall effect is a Hall voltage originating from the magnetization of a ferromagnet rather than from an external magnetic field. Karplus and Luttinger \cite{KarLut} attributed its origin to spin-orbit interaction. Because this interaction is a relativistic (hence, usually a weak) effect, and is inherent both in the Hamiltonian of the host crystal and in the potentials of impurities, there is a number of competing contributions to AHE conductivity. This explains the long history of controversies in the theory of AHE. Original theory of AHE \cite{KarLut} resulted in the transverse conductivity that was independent of the impurity concentration and was expressed completely in terms of the Bloch functions of the host crystal. From this standpoint, AHE could be considered as an {\it intrinsic} phenomenon. Soon afterwards, Smit \cite{Smit} proposed an {\it extrinsic} mechanism of AHE originating from the Mott' skew-scattering of free carriers by impurities. This mechanism is completely due to the non-Bornian part of the scattering amplitude. Remarkably, the spin-dependent part of the impurity potential is tremendously enhanced by the crystal field (by six orders of magnitude in GaAs) as compared to its magnitude in vacuum \cite{EHR05}.

In a revised theory, Luttinger \cite{Lut58} found a regular expansion of the conductivity in powers of the impurity potential $\lambda$. He concluded that the leading term in the nondiagonal (Hall) part of the conductivity is of the order of $1/(\lambda n_i)$, $n_i$ being the impurity concentration, while the next term is of the order of $(\lambda n_i)^0$, i.e., it is independent of both of $\lambda$ and $n_i$. Moreover, second term does not depend on any properties of the impurities and is completely determined by the  properties of the host crystal. The physical origin of this remarkable behavior of second term is not clear from Luttinger's calculations that are rather cumbersome. Meantime, early experimental data on $n_i$ dependence of AHE in iron quoted in Ref.~\onlinecite{Lut58}, and also some more recent data, seem to suggest the dominance of second term in some materials.  Stability of second term of the expansion in $\lambda$ and its large magnitude, if really supported by experimental data, suggest existence of some fundamental requirements protecting this stability.

A different {\it extrinsic} mechanism for AHE, i.e., caused by impurity scattering, was proposed by Berger \cite{Berger}. The point is that spin-orbit scattering is accompanied by ``side jump" of an electron in the configurational space. In systems possessing high spatial symmetry it is directed along $({\bf k}\times\mbox{\boldmath$\sigma$})$, $\bf k$ being electron quasi-momentum and {\boldmath$\sigma$} a vector of Pauli matrices. In a spin polarized system, $\langle\mbox{\boldmath$\sigma$}\rangle\neq0$, side jump produces electric current that is transverse to the driving field $\bf E$ and proportional to magnetization. This mechanism results in the same dependence of the anomalous Hall resistance on $n_i$ as the second term of the Luttinger theory \cite{Lut58}. Most remarkably, contribution of this mechanism to the anomalous Hall conductivity depends neither of $n_i$ nor on the electron mean free time $\tau$, hence, this extrinsic effect bears features typical of an intrinsic phenomenon.

More recently, the concepts of Berry connection and Berry curvature in the momentum space were applied to the theory of AHE \cite{SN99,Jung02,ON02,BB03,Hald04,Dug05,Nag06,WM06}. This approach is based on a close relation between the second term of the Luttinger theory [see Eqs.~(2.17), (3.15), and (4.21) of Ref.~\onlinecite{Lut58}] and the topological invariant of the theory of the Integer Quantum Hall Effect by Thouless {\it et al.} \cite{TKNdN82,Kom85}. It might have potentiality to explain the remarkable stability of the side jump term and to justify applying equations derived in the dilute limit to dirty materials. For Quantum SHE, such stability was proven numerically by Sheng {\it et al.} \cite{SWSH06}. 

However, to best of my knowledge, no general connection between the side jump mechanism and Berry curvature has been established so far. In this paper, I apply the equations for the side jump contribution derived by Nozi\`{e}res and Lewiner \cite{NL73} for AHE and by Engel {\it et al.} \cite{EHR05} for SHE to compare them with Berry curvature. The model used in these papers is applicable to electrons in bulk GaAs.

In the framework of ${\bf k}\cdot{\bf p}$ theory, the Hamiltonian of extrinsic spin-orbit coupling for electrons in GaAs type semiconductors is
\be
H_{\rm so}=\lambda(\mbox{\boldmath$\sigma$}\times{\bf k})\cdot\nabla V({\bf r})\,,
\label{eq1}
\ee
where the potential energy $V({\bf r})$ is a smooth function of $\bf r$. Applying (\ref{eq1}) to a homogeneous external field $\bf E$, with $V({\bf r})=-e{\bf E}\cdot{\bf r}$ (for electrons, $e<0$), we get
\be
H_{\rm so}^E=-e\lambda(\mbox{\boldmath$\sigma$}\times{\bf k})\cdot{\bf E}\equiv -e{\bf E}\cdot{\bf r}_{\rm so}({\bf k})\,.
\label{eq2}
\ee
Eq.~(\ref{eq2}) indicates existence of spin-orbit contribution
\be
{\bf r}_{\rm so}({\bf k})=\lambda(\mbox{\boldmath$\sigma$}\times{\bf k})
\label{eq3}
\ee
to the operator $\hat{\bf r}$ of the electron coordinate
\be
\hat{\bf r}={\bf r}+{\bf r}_{\rm so}({\bf k})\,.
\label{eq4}
\ee
While Eq.~(\ref{eq1}) reminds the Darwin term in Pauli equation, the coefficint $\lambda$ is strongly enhanced compared to its vacuum value. In narrow gap semiconductors, in the framework of $8\times8$ Kane model, $\lambda=(P^2/3)[1/E_G^2-1/(E_G+\Delta)^2]$, where $E_G$ is the forbidden gap, $\Delta$ is the spin-orbit splitting of valence bands, and $P$ is a properly normalized interband momentum matrix element \cite{Yafet}.

Semiclassical arguments based on electron dynamics in the field of an impurity center that I do not reproduce here result in a spin dependent side jump. In turn, side jump results in a transverse Hall current \cite{NL73}
\be
{\bf J}_{\rm sj}=-2n\lambda{{e^2}\over{\hbar}}(\langle\mbox{\boldmath$\sigma$}\rangle\times{\bf E})\,,
\label{eq5}
\ee
where $n$ is the electron concentration. This current originates from the anomalous coordinate ${\bf r}_{\rm so}$ of Eq.~(\ref{eq3}). A similar equation for the spin Hall current under the SHE conditions was derived in Ref.~\onlinecite{EHR05}. Remarkably, ${\bf J}_{\rm sj}$ is independent of any specific properties of the scatterers that produced the current.

Eq.~(\ref{eq4}) suggests that the Hamiltonian of a perfect crystal in a homogeneous electric field $\bf E$ can be written as
\be
H=H_0({\bf k})-e{\bf E}\cdot\hat{\bf r}\,.
\label{eq6}
\ee
For our purposes, it is enough to choose for $H_0({\bf k})$ the nonrelativistic part of the ${\bf k}\cdot{\bf p}$ Hamiltonian. Then, the relativistic part ${\bf v}_{\rm so}$ of the velocity operator ${\bf v}=(i/\hbar)[H,\hat{\bf r}]$ equals
\bea
{\bf v}_{\rm so}&=&-i{e\over\hbar}\lambda\{[(\mbox{\boldmath$\sigma$}\times{\bf k})\cdot{\bf E},{\bf r}]+[({\bf E}\cdot{\bf r}),(\mbox{\boldmath$\sigma$}\times{\bf k})]\}\nonumber\\
&=&2{e\over\hbar}\lambda(\mbox{\boldmath$\sigma$}\times{\bf E})\,.
\label{eq7}
\eea
Because ${\dot{\bf k}}=(i/\hbar)[H,{\bf k}]=e{\bf E}/\hbar$, it follows from (\ref{eq3}) that
\be
{\dot{\bf r}_{\rm so}}={e\over\hbar}\lambda(\mbox{\boldmath$\sigma$}\times{\bf E})\,.
\label{eq8}
\ee
Therefore
\be
{\bf v}_{\rm so}=2{\dot{\bf r}_{\rm so}}\,.
\label{eq9}
\ee
Factor of 2 in (\ref{eq7}) and (\ref{eq9}) originates from the field term in the Hamiltonian of Eq.~(\ref{eq6}). An expression for the intrinsic contribution to the anomalous Hall current immediately follows from (\ref{eq7}) 
\be
{\bf J}_{\rm int}=2n\lambda{{e^2}\over{\hbar}}(\langle\mbox{\boldmath$\sigma$}\rangle\times{\bf E})\,.
\label{eq10}
\ee
Therefore, for the Hamiltonian $H_{\rm so}$ of Eq.~(\ref{eq1}) the current ${\bf J}_{\rm sj}$ has exactly the same magnitude but the opposite sign to ${\bf J}_{\rm int}$, ${\bf J}_{\rm sj}=-{\bf J}_{\rm int}$. It is important to emphasize that while the derivation of (\ref{eq10}) from Eq.~(\ref{eq6}) is straightforward and formally correct, the Hamiltonian $H$ is faulty because it does not support any stationary state. For this reason, ${\bf J}_{\rm int}$ can contribute to the physical anomalous Hall current only in conjunction with the terms originating from impurity scattering. Recent experience with the theory of SHE demonstrates convincingly that impurity scattering influences the results dramatically \cite{ERH07}.

Now let us express ${\bf r}_{\rm so}$ and ${\bf v}_{\rm so}$ in terms of microscopic theory. For ${\bf r}_{\rm so}$, a standard expression of the ${\bf k}\cdot{\bf p}$ theory can be applied
\be
{\bf r}_{\rm so}({\bf k})=i\int {\bar u}_{\bf k}({\bf r})\nabla_{\bf k}u_{\bf k}({\bf r})~d{\bf r}\,,
\label{eq11}
\ee
where $u_{\bf k}({\bf r})$ are two-component Bloch spinors (the periodic part of Bloch eigenfunctions), and integration is performed over a unit cell. The right hand side of (\ref{eq11}) is usually termed as Berry connection; sometimes the opposite sign convention is applied. Then, by definition, Berry curvature is
\be
{\bf F}({\bf k})=\nabla_{\bf k}\times{\bf r}_{\rm so}=i\int \nabla_{\bf k}{\bar u}_{\bf k}({\bf r})\times\nabla_{\bf k}u_{\bf k}({\bf r})~d{\bf r}\,.
\label{eq12}
\ee
It follows immediately from (\ref{eq12}) that for the anomalous coordinate ${\bf r}_{\rm so}$ of Eq.~(\ref{eq3})
\be
{\bf F}({\bf k})=\lambda~\nabla_{\bf k}\times(\mbox{\boldmath$\sigma$}\times{\bf k})=2\lambda\mbox{\boldmath$\sigma$}\,.
\label{eq13}
\ee
Here the factor of 2 comes from the double cross product because $\epsilon_{ij\ell}\epsilon_{mj\ell}=2\delta_{im}$, $\epsilon_{ij\ell}$ is a Levi-Civita antisymmetric tensor. Comparing with (\ref{eq7}) results in the equation
\be
{\bf v}_{\rm so}={e\over\hbar}({\bf F}\times{\bf E})\,,
\label{eq14}
\ee
relating anomalous velocity ${\bf v}_{\rm so}$ to the Berry curvature $\bf F$. Therefore, the mean value of the curvature $\langle{\bf F}\rangle=2\lambda\langle\mbox{\boldmath$\sigma$}\rangle$ can be used for calculating ${\bf J}_{\rm int}$ of Eq.~(\ref{eq10}) as
\be
{\bf J}_{\rm int}=n{{e^2}\over{\hbar}}(\langle{\bf F}\rangle\times{\bf E})\,.
\label{eq15}
\ee
For a parabolic spectrum, $H_0=\hbar^2{\bf k}^2/2m$, a dynamic equation for the electron coordinate is
\be
{{d}\over{dt}}{\hat{\bf r}}={i\over\hbar}[H,{\hat{\bf r}}]={\bf v}({\bf k})+{{e}\over{2\hbar}}({\bf F}\times{\bf E})\,,
\label{eq16}
\ee
with ${\bf v}({\bf k})=\hbar {\bf k}/m$. The coefficient $1/2$ in second term of (\ref{eq16}) has the same origin as the factor of 2 in Eq.~(\ref{eq9}).

Therefore, for the spin-orbit Hamiltonian of Eq.~(\ref{eq1}), the side jump contribution ${\bf J}_{\rm sj}$ to anomalous Hall current can be expressed in terms of the Berry curvature ${\bf F}({\bf k})$, and ${\bf J}_{\rm sj}=-{\bf J}_{\rm int}$. 
Equal magnitude and mutual cancelation of a number of contributions to the anomalous Hall current is a well known fact \cite{NL73}; more recently, it was discussed in Ref.~\onlinecite{SNSN05} in terms of a semiclassical theory. However, it is still not understood which of these cancelations are accidental and which follow from general requirements.

One general comment should be made regarding Eq.~(\ref{eq11}) that was written for noncentrosymmetric crystals with a lifted spin degeneracy. For centrosymmetric crystals $u_{\bf k}$ should be substituted by $u_{\alpha\bf k}$ with $\alpha$ playing a role of a spin index. Then ${\bf r}_{\rm so}$ becomes a $2\times2$ matrix in spin space whose spatial components $x_{\rm so}^j$ do not transform as the components of a vector of the configurational space under the rotations in the spin space, and this is valid also for the components of its curl, $\nabla_{\bf k}\times{\bf r}_{\rm so}$, i.e., for the Berry curvature. Blount has shown \cite{Blount} that to restore the correct transformation properties of $\nabla_{\bf k}\times{\bf r}_{\rm so}$, it should be redefined as
\be
{\bf F}({\bf k})=\nabla_{\bf k}\times{\bf r}_{\rm so}({\bf k})-i{\bf r}_{\rm so}({\bf k})\times{\bf r}_{\rm so}({\bf k})\,,
\label{eq17}
\ee 
what is equivalent to a redefinition of Berry curvature. As applied to ${\bf r}_{\rm so}$ of Eq.~(\ref{eq3}), this is equivalent to adding a term $2\lambda^2(\mbox{\boldmath$\sigma$}\times{\bf k})\cdot{\bf k}$ to the right hand side of Eq.~(\ref{eq13}). This redefinition has no physical consequences for the quantities that depend on ${\bf F}({\bf k})$ only through its trace over spin indices. Indeed, for the projection of ${\bf r}_{\rm so}({\bf k})\times{\bf r}_{\rm so}({\bf k})$ onto arbitrary direction $\hat{\bf m}$,
\be
{\rm tr}\{({\bf r}_{\rm so}\times{\bf r}_{\rm so})\cdot{\hat{\bf m}}\}=\epsilon_{ij\ell}~m^\ell{\rm tr}\{x_{\rm so}^ix_{\rm so}^j\}=0
\label{eq18}
\ee
because ${\rm tr}\{x_{\rm so}^ix_{\rm so}^j\}={\rm tr}\{x_{\rm so}^jx_{\rm so}^i\}$. Therefore, after averaging over spin states at any given $\bf k$, second term in (\ref{eq17}) vanishes. Also, this term can be sometimes disregarded because it is of the second order in the spin-orbit coupling constant.

I am grateful to H.-A. Engel and B. I. Halperin for inspiring collaboration on papers Ref.~\onlinecite{ERH07} and \onlinecite{EHR05}, and to N. Nagaosa for a fruitful discussion. 


\end{document}